\documentclass[runningheads]{article}
\usepackage[english]{babel}
\usepackage{amsmath}
\usepackage{amsfonts}
\usepackage{amssymb}
 \usepackage{epsfig}
 \usepackage{graphicx}%
 \usepackage{verbatim}
\begin{document}

\title{Are your data really Pareto distributed?}

\maketitle

\begin{abstract}
Pareto distributions, and power laws in general, have demonstrated to be very useful models to describe very different phenomena, from physics to finance. In recent years, the econophysical literature has proposed a large amount of papers and models justifying the presence of power laws in economic data.\\
Most of the times, this Paretianity is  inferred from the observation of some plots, such as the Zipf plot and the mean excess plot. If the Zipf plot looks almost linear, then everything is ok and the parameters of the Pareto distribution are estimated. Often with OLS.\\
Unfortunately, as we show in this paper, these heuristic graphical tools are not reliable. To be more exact, we show that only a combination of plots can give some degree of confidence about the real presence of Paretianity in the data.\\
We start by reviewing some of the most important plots, discussing their points of strength and weakness, and then we propose some additional tools that can be used to refine the analysis.
\end{abstract} 

\section{Introduction}

This is not a paper about estimation. We are not going to discuss estimation methods for Paretian distributions or power laws, stating if it is better to use OLS, MLE, Hill-type estimators or minimization algorithms on goodness-of-fit statistics. A series of recent good papers and books on the subject is available in the literature, e.g. \cite{Cihu, Clau, Embr, Falk, Gaba, Gaib}, and we refer the reader to them.\\
This paper deals with a somehow surprisingly neglected step in the study of Paretianity in empirical data, i.e. the verification of the power law hypothesis. Estimating the parameters of a power law, namely the lower bound and the shape/tail coefficient, is indeed meaningful only if the used data are actually drawn by some Paretian distribution. Conversely, if the observations in the sample are distributed according to other distributions, the estimation of the Pareto parameters does not make much sense. Or more, it may be a dangerous waste of time. \\
In the literature there are many different methods to test the power law hypothesis, from goodness-of-fit tests, in particular the Kolmogorov-Smirnov \cite{Clau} and the Anderson-Darling \cite{Cihu}, to more immediate graphical tools. It goes without saying that plots are definitely the most used, and sometimes abused, instruments. The reason is simple: the different available plots are essentially immediate graphical tests on some fundamental properties of power laws. They are easy to produce, and they do not require entering into more ``complicated" statistical tests; all this makes them very attractive for practitioners and all those researchers not interested in statistics itself. \\
However, and this is what we aim to address in the paper, graphical tools are just heuristic tools, and their interpretation is not as easy and straightforward as it may seem. We are going to show that, quite often, graphical tools can lead to wrong decisions, especially if one relies on just one type of plots. \\
In what follows we will focus our attention on some of the most used plots, such as the Zipf and the mean excess function plots. For each plot we will try to give guidelines for its correct interpretation, always stressing that only the combination of different tools can give a good idea about the nature of the analyzed data. Basic codes for producing the plots are also given in the appendix.\\ 
In the second part of the paper we also discuss two additional plots, which are not used in the literature, especially in the econophysical one, but which could represent useful instruments for identifying Paretianity.\\
\\
Anyway, before entering into the core discussion of the paper, let us refresh some basic facts about Pareto distributions and power laws.\\
A random variable $X$ is said to follow a Pareto distribution if its density function $f(x)$ is such that
\begin{equation}
f(x)=\frac{\alpha x_0^\alpha}{x^{\alpha+1}}, \qquad 0<x_0\leq x,
\end{equation}
where $\alpha$ is the so-called shape parameter, which measures the heaviness of the right tail, and $x_0$ is a scale parameter. The corresponding cumulative distribution function (cdf)  is thus 
\begin{equation}\label{pare}
F(x)=1-\left(\frac{x}{x_0} \right)^{-\alpha}, \qquad 0<x_0\leq x.
\end{equation}
The parameter $\alpha$ is definitely the most important quantity for a Pareto distribution, since it determines its behavior. For example, the $k$-th moment of a Pareto random variable exists only for $k<\alpha$, and it is equal to
\begin{equation}
E[X^k]=\frac{\alpha x_0^k}{\alpha-k}.
\end{equation}
The smaller $\alpha$, the fatter the right tail of the distribution. For $\alpha<2$ the Pareto distribution has an infinite variance. For $\alpha<1$ the expected value does not exist.\\
The Pareto distribution was introduced by Vilfredo Pareto, an Italian economist and engineer, in \cite{Par}. It represents one of the most famous continuous distributions and it is widely used in economics, finance, econophysics and natural sciences.\\
To be more precise, the distribution we have just introduced is known as the Pareto I, and its classical notation is $Par(x_0,\alpha)$. Over the years, starting from Pareto himself, many generalizations have been proposed. A simple one is the Pareto II, also known as Lomax distribution, where
\begin{equation}
F(x)=1-\left[1+\frac{x}{b} \right]^{-\alpha}, \qquad x>0.
\end{equation}
It is worth underlining that, if $X \sim Par_{II}(b,\alpha)$, then $X+b\sim Par(b,\alpha)$.\\
Another very famous generalization is the GPD, or Generalized Pareto distribution, very important in extreme value theory, for which
\begin{equation}
F(x)=\begin{cases}1-\left(1+\frac{\xi(x-\nu)}{\beta} \right)^{-\frac{1}{\xi}} & \xi \ne 0\\
1-\exp\left(-\frac{x-\nu}{\beta} \right) & \xi=0
\end{cases},
\end{equation}
where $x\geq \nu$ for $\xi\geq 0$, $\nu\leq x \leq \nu-\beta/\xi$ for $\xi<0$, $\nu,\xi \in \mathbb{R}$ and $\sigma>0$. Notice that the a GPD exactly corresponds to a Pareto I with $\alpha=1/\xi$ when $\xi>0$ and $\nu=\beta/\xi$.\\
More in general, Pareto distributions can be seen as power laws, i.e. distributions for which
\begin{equation}
f(x)\propto L(x) x^{-\alpha},
\end{equation}
where $L(x)$ is a slowly varying function ($\lim_{x \to \infty}\frac{L(cx)}{L(x)}=1$, with $c>0$ constant; for more details see \cite{Embr}). It is easy to verify that the Pareto I is nothing more than a general power law where $L(x)$ is a constant incorporating $\alpha$. \\
A rather complete taxonomy of Pareto distributions and power laws is available in \cite{Klei} and \cite{Joko}, and we refer the reader to them.\\
In what follows, also considering the several different specifications available in the empirical literature (once again see \cite{Klei}), we do not make any distinction about the possible Pareto distributions. This is due to the fact that all the plots we discuss and present do work in general for power laws. Hence, from now on, when we speak about the \textit{Paretianity hypothesis}, we simply mean that our data come from a power law. This law can be a pure Pareto I, a GPD, but also a more general representation with a slowly varying component. 

\begin{figure}
  \centering
    \includegraphics[scale=0.7]{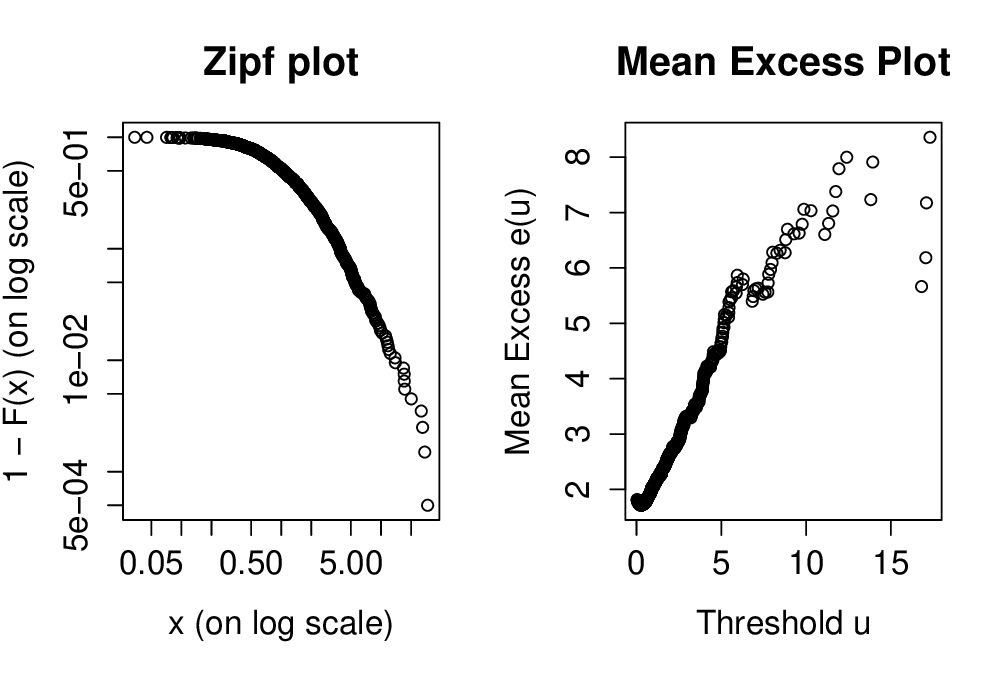}
  \caption{Examples of Zipf plot and Mean Excess Plot.}
    \label{fig1}
\end{figure}

\section{The Zipf plot}
\begin{figure}
  \centering
    \includegraphics[scale=0.5]{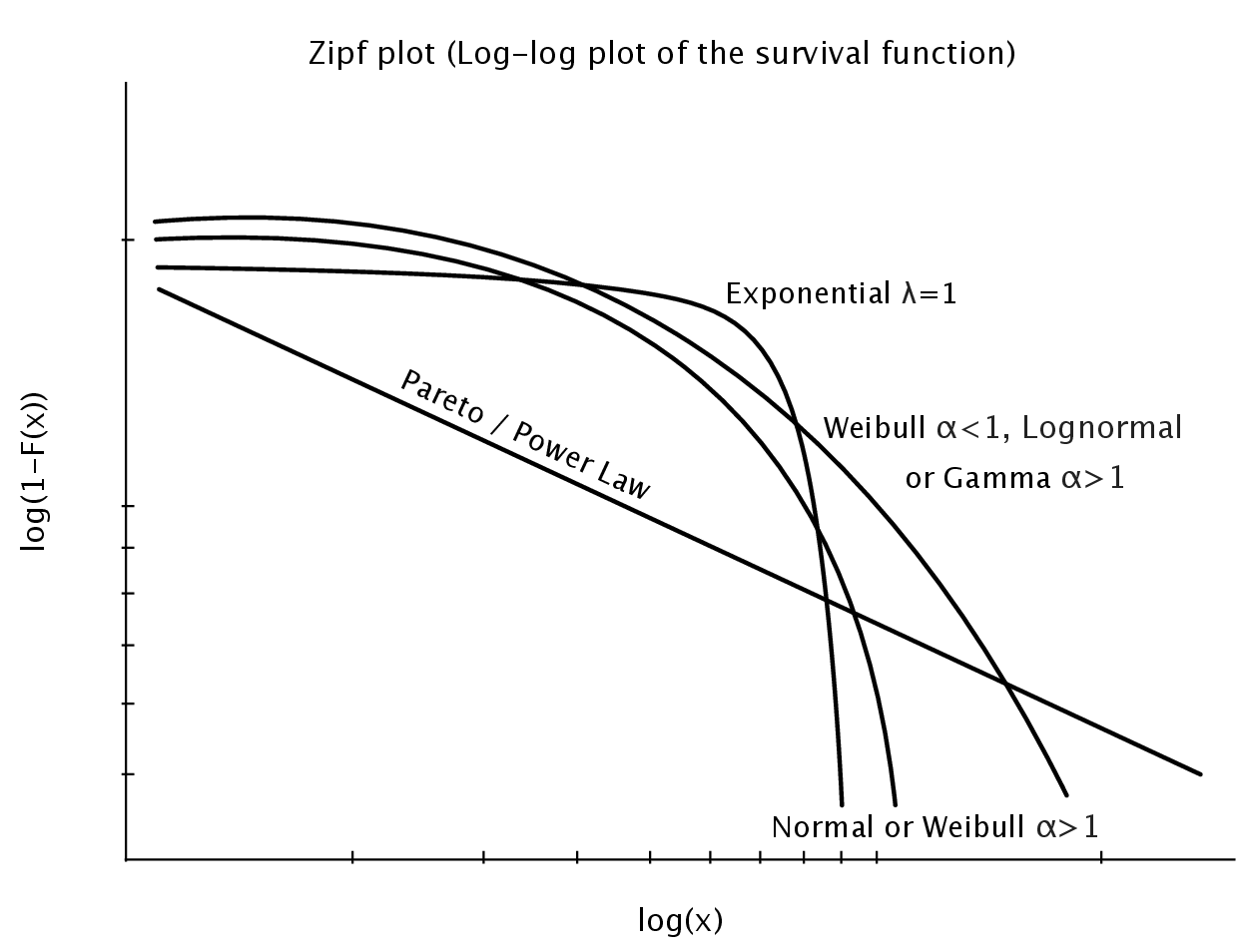}
  \caption{Zipf plot behavior for some classical distributions.}
    \label{emplot}
\end{figure}

The Zipf plot is probably the most used and abused plot for verifying the presence of Paretianity in the data. The original plot was proposed in \cite{Zipf} and it was constructed on binned observations. Here we present a different version based on the empirical survival function. However, later in the paper, we also discuss the use of binning.\\
Consider a standard Pareto I distribution, whose cdf is given in equation (\ref{pare}). The survival function $\bar{F}(x)=1-F(x)$ is thus equal to
\begin{equation}\label{survi}
\bar{F}(x)=\left(\frac{x}{x_0} \right)^{-\alpha}, \qquad 0<x_0\leq x.
\end{equation}
Now, let us take the logs on both sides of equation (\ref{survi}), getting $\log(\bar{F}(x))=\alpha \log (x_0) -\alpha \log(x)$. By substituting $C=\alpha \log (x_0)$, we get $\log(\bar{F}(x))=C -\alpha \log(x)$, i.e. a negative linear relationship between the logarithm of the survival function and the logarithm of $x$. The slope of the line is equal to $-\alpha$. This derivation holds for a Pareto I, but it is easy to obtain similar results for all Paretian/power law distributions, up to rescaling and changes of variable.\\
We now have all the ingredients to create a Zipf plot, i.e. a plot in which the logarithm of the empirical survival function is plotted against the logs of the ordered values of $x$. If the data follow a power law, we expect to observe a more or less negative linear relationship in the graph. On the left side of Figure \ref{fig1} an example is given.\\
Figure \ref{fig1} allows us to discuss a little more about the Zipf plot. Naturally a single straight line can only be observed for purely Paretian data, but generally this is not the case. In most empirical analyses, where some Paretian behavior is present, the Paretianity accounts for a certain amount of the data, in particular the upper tail of the distribution. In Figure \ref{fig1} we can observe that the Zipf plot starts as a curve, and that a linear behavior is only observable for $x>2$. For these values, our plot suggests the possible presence of  a Paretian tail. \\
From a heuristic point of view, the Zipf plot can thus be used to identify the threshold value above which Paretianity seems to hold. That value $x_0$ will simply be the one above which the Zipf plot shows a negative linear behavior.\\
The Zipf plot can also be used to heuristically check alternative distributional hypotheses. In Figure \ref{emplot} the theoretical behavior of the Zipf plot for some famous distributions is given. The cases there presented account for ``pure" distributions. The interpretation of a Zipf plot in case of mixtures is much more difficult. While it is not problematic to distinguish between an exponential and a Pareto, it may be more dangerous to discriminate between a Normal and a Lognormal, or a Lognormal and a Pareto, just on the basis of a Zipf plot. Especially for lognormal data, it must be stressed that the right tail tends to open on the right hand side, sometimes looking quasi-linear, for large values of $\sigma$. It is in fact known that the heavy-tailed nature of the lognormal distribution reveals itself for lognormal data with extreme variability \cite{Embr} \cite{Klei}.\\

\subsection{Binning}
Depending on the type and quality of the analyzed data, the Zipf plot may present a rather noisy behavior in the upper tail of the distribution. \\
As well explained in \cite{Calda}, this is essentially due to the fact that, in a sample of size $N$, with range 10-10000, a large number of observations are likely to fall in the interval $10-100$, while just a smaller amount of data will fall in the interval $1000-10000$. However, on a logarithmic scale, these two intervals have the same size, and since $N$ is necessarily finite, it is hard to avoid some noise in the upper tail of the Zipf plot.\\
A possible way of reducing the noise in the plot is to use binning. The basic idea is to choose the right intervals, i.e. bins (as when plotting a histogram), and to average the observations within each bin, in order to reduce noise, given the fact that the sum of fluctuations from the average is equal to zero for statistical noise.\\
In practice, after sorting the observations from the smallest to the largest, one divides the $x$-axis into a certain number of bins, say $B$. For each bin $b=1,...,B$, one takes middle point $\bar{x}_b$ (the average of the bin's endpoints), and $\bar{y}_b$ as the average of all the $y$'s corresponding to the $x$'s falling in $b$.\\
The interesting feature of binning when dealing with power laws is that one can use logarithmic bins. The first step is to choose the size $s>1$ of the first bin; $s$ will represent the basis of the logarithmic progression of bins $(s^1,s^2,...,s^B)$. For simplicity, let's assume $s=2$. Then the second bin will have size equal to $2^2=4$ and the third $2^3=8$, and so on. This procedure guarantees that the bins are equally spaced in the logs, but coming back to the original non-log data, we are considering bins of steeply increasing size, thus trying to have more or less the same number of observations within each bin.\\ 
Once the bins $b=1,..,B$ have been created, one can take $\bar{x}_b$ to be the geometric mean of the two endpoints of bin $b$, and $\bar{y}_b$ to be the arithmetic average of all the $y$'s corresponding to $b$. Once $\bar{x}_b$ and $\bar{y}_b$ have been computed for all $b=1,...,B$, one can plot them in the Zipf plot in search for power law behavior\footnote{Statistical programs like R and Matlab provide useful functions to create log bins and to perform all the operations we have described, see for example the $hist$ function in R.}. \\
A big issue in binning, and this is particularly true for logarithmic bins, is how to choose the optimal size/basis for the bins. The answer is simple, yet annoying: through a trial-and-error approach and experience. The trade-off between noise reduction and information loss is in fact evident. Especially for small data sets, the choice of too large bins will cause the loss of worth-investigating behaviors in the tail of the plot. Conversely, too small bins may not be able to sufficiently reduce noise.\\
For a less heuristic approach, one could work with the cumulative distribution function or the hazard rate. For more details, please refer to \cite{Calda} and \cite{Klei}.

\section{The mean excess function plot (Meplot)}
As the name suggests, the mean excess function plot is based on the behavior of the mean excess function. It is a plot largely used in extreme value theory \cite{Embr}, but less popular in the econophysical literature \cite{Cihu}.\\
Let $X$ be a random variable with distribution $F$ and right endpoint $x_F$ (i.e. $x_F=\sup\{x \in \mathbb{R}: F(x)<1\}$). The function
\begin{equation}
e(u)=E[X-u|X>u]=\frac{\int_u^\infty (t-u) \text{d}F(t)}{\int_u^\infty \text{d}F(t)}, \qquad 0<u<x_F,
\end{equation}
is called mean excess function of $X$.\\
From an empirical point of view, the ME of a sample $X_{1}$, $X_{2}$,..., $X_{n}$ is easily computed as
\begin{equation}
e_{n}(u)=\frac{\sum_{i=1}^{n}(X_{i}-u)}{\sum_{i=1}^{n}1_{\{X_{i}>u\}}},
\end{equation}
that is the sum of the exceedances over the threshold $u$ divided by the number of data points exceeding $u$.\\
Together with hazard rates, the mean excess function represents a fundamental tool of insurance mathematics \cite{Embr}.\\
Interestingly, the ME is a way of characterizing distributions within the class of continuous distributions \cite{Klei}, and this fact can be used to check the Paretianity hypothesis in the data. The Pareto distribution (and its generalizations) is indeed the only distribution characterized by the so-called van der Wijk's law \cite{vanWijk}. This law, which was originally stated in the field of income and wealth studies, asserts that the average income of all the people above a given level $u$ is proportional to $u$ itself, i.e.
\begin{equation} \label{meq}
\frac{\int_u^\infty tf(t) \text{d}t}{\int_u^\infty f(t)\text{d}t}=c u, \qquad c>0.
\end{equation}
Clearly the left hand side of equation (\ref{meq}) is the mean excess function. In other terms, the Pareto distribution is characterized by a mean excess function that is linear in the threshold $u$. To be more exact, the ME of a standard Pareto I distribution is equal to
\begin{equation}
e_{PA_I}(u)=\frac{u}{\alpha-1},\qquad \alpha>1,
\end{equation}
so that $c=(\alpha-1)^{-1}$. The van der Wijk's law hence does hold. \\
This linearity also holds for more general definitions of Pareto distribution, including the Pareto II (or Lomax), the GPD and power laws. For example, a Pareto II has
\begin{equation}
e_{PA_{II}}(u)=\frac{u+b}{\alpha-1},\qquad \alpha>1,
\end{equation}
and a GPD
\begin{equation}
e_{GPD}(u)=\frac{\beta+\xi u}{1-\xi},\qquad \beta+\xi u>0.
\end{equation}
For power laws, especially if they have a slowly-varying component, we can have a slightly different behavior, and the linearity can only be approximated. For instance, in the log-gamma case, where $f(x)=\frac{\alpha^\beta}{\Gamma(\beta)}(\log x)^{\beta-1}x^{-\alpha-1}$, $\alpha,\beta>0$, we have
 \begin{equation}
e_{LG}(u)=\frac{u}{\alpha-1}(1+o(1)),\qquad \alpha>1.
\end{equation}

\begin{figure}
  \centering
    \includegraphics[scale=0.5]{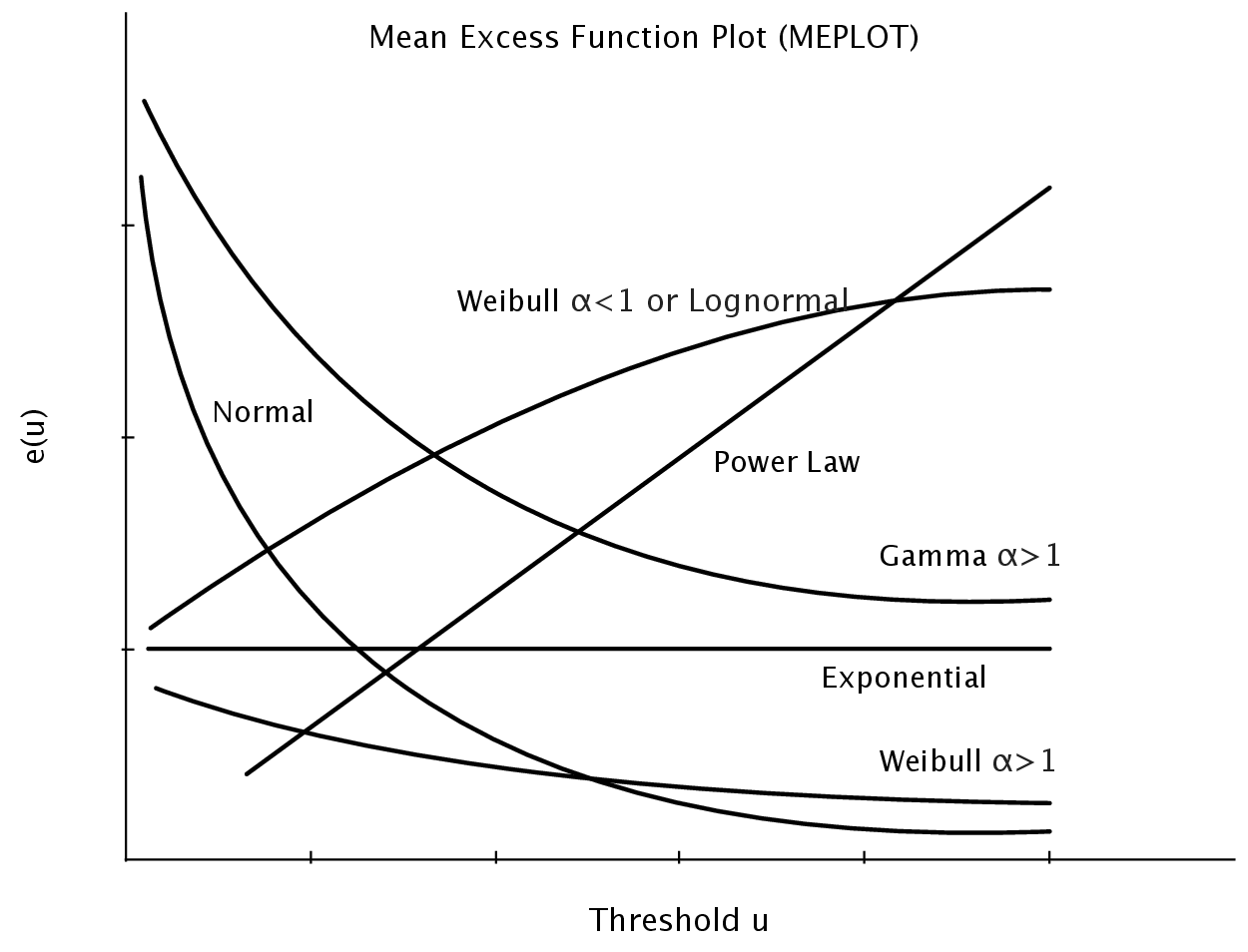}
  \caption{Shape of the mean excess function $e(u)$ for some classical distributions as a function of the threshold $u$.}
  \label{meplot}
\end{figure}

Similar results do hold for the Burr (Singh-Maddala) and other Paretian distributions, for which we refer to \cite{Klei} and \cite{Joko}.\\
Hence, if we create a graph, in which the points $\{(X_{i:n},e_n(X_{i:n})): i=1,...,n\}$ are plotted, $X_{1:n},X_{2:n},...,X_{n:n}$ being the order statistics of our data set, what we obtain is called mean excess (function) plot, or MEPLOT.\\
Given the properties of the mean excess function for Paretian random variables, a meplot showing a linearly increasing trend can be considered a signal of Paretianity in the data. An example of mean excess plot is given in Figure \ref{fig1}, on the right.\\
Naturally, the meplot is an empirical tool, hence there are some important things we need to keep in mind when reading it:
\begin{itemize}
\item An upward linear trend is a signal of Paretianity, but it is not possible to discriminate within the Paretian family;
\item The mean excess function is extremely sensitive to changes in the data, especially for the very large observations. This is due to the fact that, for large thresholds $u$, the corresponding $e_n(u)$ may depend on just a few observations. Typically, this problem is solved by not considering the largest values of $e_n(u)$, i.e. by ignoring its behavior for the largest 5-10 threshold values \cite{Embr}. In extreme value theory, this can sometimes be a further problem, given the limited number of observations, but it is not the case in most econophysical applications.
\end{itemize}
The meplot is a rather powerful plot, since it allows to verify the Paretian hypothesis, but also to look for alternatives. When studying size distributions, e.g. for firms' size, this is certainly a plus. \\
Figure \ref{meplot} gives instructions on how to read a meplot, by showing the behavior of $e(u)$ for different distributions. Notice that fat-tailed distributions, such as the Pareto and, especially for large values of $\sigma$, the lognormal, typically show a ME tending to infinity. In case of lognormally distributed random variables, we have
\begin{equation}
e_{LN}(u)=\frac{\sigma^2 u}{\log(u)-\mu}(1+o(1)).
\end{equation}
Other distributions, such as for example the exponential $Exp(\lambda)$, have a totally different behavior, with $e_{EXP}(u)=\lambda^{-1}$.\\
In case of mixture of distributions with Paretian tails, the meplot can also represent a heuristic way of identifying the threshold $u$, above which Paretianity holds. The idea is simply to look for the value of $u$ above which the empirical mean excess function $e_n(u)$ looks almost linear and increasing.

\section{From theory to practice}
\begin{figure}
  \centering
    \includegraphics[scale=0.7]{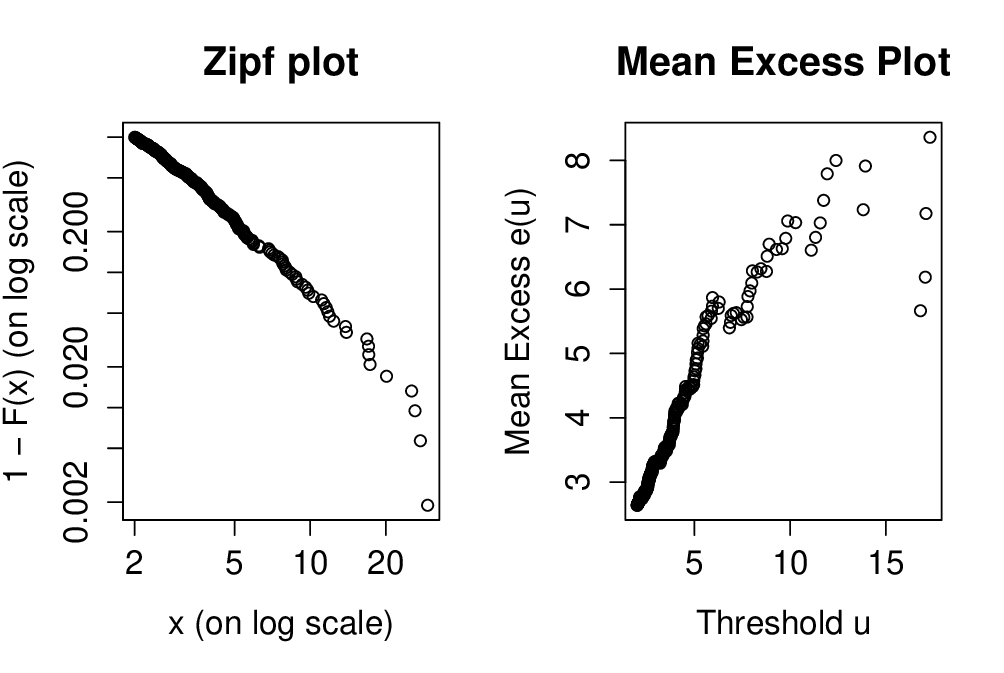}
  \caption{Zipf and mean excess plot for the same data of Figure 1, but focusing on the observations greater than 2.}
    \label{troncata}
\end{figure}

Let us come back to Figure \ref{fig1}, where the Zipf plot and the meplot of an empirical data set with 500 observations are given.\\
Looking at the Zipf plot, we can clearly see that the data do not come from a purely Paretian distribution. In fact, the log-log plot of the survival function is not entirely linear, but it also show a curvature on the left hand side. However, we can easily observe a linear behavior with negative slope in the right part of the plot, especially for values of $x$ greater than 2.\\
If we now consider the meplot in Figure \ref{fig1}, we arrive to the same conclusions: some Paretianity definitely seems to be present. For $u>2$, the mean excess function shows indeed a clearly upward trend, while for $u \leq 2$ a first decreasing and then constant behavior is observable (especially if we zoom in). \\
Figure \ref{troncata} is obtained by just focusing our attention on the observations greeter than 2. Now, instead of 500 observation we are left with 118 data points, i.e. the top 23.6\%. The Zipf plot and the mean excess plot are clearly as we would expect in case of Pareto distributed data, linearly decreasing and linearly increasing respectively. The mean excess plot shows some volatility for the greater values of $u$, but how we have said, this is a rather standard behavior.\\
Hence, looking at these plots, we can come to the conclusion that our data are drawn from a (mixture) distribution showing a clearly Paretian upper tail. Since this tail accounts for more than 20\% of all the observations, the Paretian behavior is rather important.\\
Unfortunately there is a problem: we can guarantee that the data in Figures \ref{fig1} and \ref{troncata} are not Paretian: they are randomly generated from a lognormal distribution. Bad news.\\
At this point, the reader could argue that this is not a big problem. At the end of the day, the lognormal distribution can be a definitely heavy-tailed distribution for large values of $\sigma$, as shown in \cite{Embr} and \cite{Falk}. For large $\sigma$, a lognormal distribution may possess such a fat right tail that the two plots are not able to distinguish between, say, a $Par(x_0,2.5)$ and a $lognormal(\mu, 20)$. Since it is evident that, with actual data, it is difficult to obtain the perfect theoretical curves of Figures \ref{emplot} and \ref{meplot}, it may also be difficult to discriminate among very fat-tailed distributions. In other words, given the data, both models could be considered satisfactory; what is relevant is the presence of a fat-tail on the right hand side.\\
But unfortunately there is another problem. Those data in Figure \ref{fig1} and \ref{troncata} are sampled from a lognormal distribution with mean $\mu=0$ and variance $\sigma^2=1$, not at all a fat-tailed distribution\footnote{The data have been generated with R and the basic $rlnorm(500,0,1)$ function.}. Very bad news! How can it be?!\\
The answer is complex, and it can be summarized as follows:
\begin{itemize}
\item For what concerns the Zipf plot of Figure \ref{fig1}, the problem is in our eyes. Since we are inclined to look for Paretianity, we are very happy to see it everywhere, even if the plot is perfectly consistent with a Lognormal distribution, as shown in Figure \ref{emplot}.
\item The Zipf plot of Figure \ref{troncata} is then a simple re-scaling of the first one, and this exacerbates our initial error.
\item The misunderstanding generated by the meplot is more subtle. The problem is in the number of observations. As shown in Figure \ref{meplot}, the lognormal distribution shows an increasing mean excess function, as the Pareto one. The main difference is that the Paretian $e(u)$ grows linearly, while the lognormal ME draws a concave curve. Unfortunately, especially for small values of $\sigma$, the lognormal mean excess function needs a lot of observations in order to show its truly concave behavior. With ``just" 500 observations we essentially observe the first part of the curve, which is quite well approximated by a linear upward line. Empirical investigations and simulations show that, on average, we need more than 10000 observations in order to clearly distinguish between a Paretian and a lognormal mean excess function. \\
A very nice treatment of the problems of the mean excess function as a tool for checking the GPD hypothesis in extreme value theory is given in \cite{Ghre}.
\item In both plots, the range of variation of our data is 0-30 (2-30 for the truncated versions). Such a small range is not really compatible with a distribution belonging to the Paretian family, which typically accounts for a larger volatility.
\end{itemize}
We have thus shown that the Zipf plot and the meplot are not sufficient to determine whether our data are Pareto distributed or not. This problem may be irrelevant if the data are really heavy-tailed, and we cannot (or we are not interested to) distinguish among compatible models. But in the simple example we have given, the standard lognormal distribution is certainly not a fat-tailed one, hence looking for Paretianity is an error. It is for instance sufficient to think about the empirical verification of Pareto and Gibrat laws in industrial dynamics \cite{Klei}, to understand the consequences of a wrong choice. We suspect that many inferences and conclusions available in the literature should be double checked, also considering that most of them simply rely on the Zipf plot.\\
However, there are good news: even if the Zipf plot and the meplot are very often unable to distinguish between, say, Pareto and lognormal random variables (or other heavy-tailed distributions), they are surely capable of rejecting the Paretian hypothesis. In fact, both the negative linear trend in the Zipf plot and the upward linear trend in the meplot are necessary conditions for the presence of Paretianity in the data. If these behaviors are not observed, then we can reject the Paretian hypothesis with confidence.\\
In the next sections, we present additional graphical tools that can be used to verify the Paretian hypothesis, thus supporting the results given by the Zipf and the mean excess function plots.

\section{The Discriminant Moment-ratio Plot}

\begin{figure}
  \centering
    \includegraphics[scale=0.5]{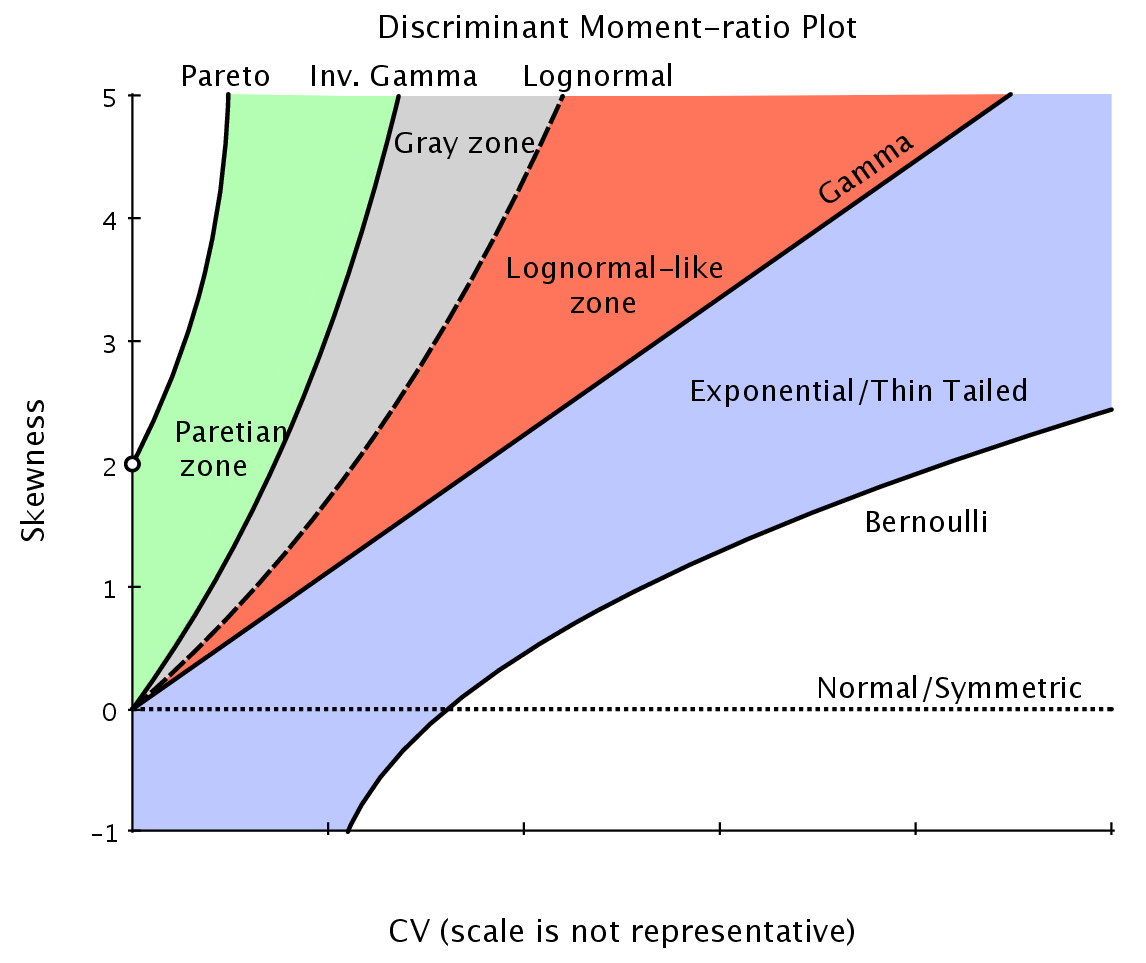}
  \caption{Guidelines for the interpretation of the Discriminant Moment-ratio Plot. Please notice that the scale of the CV axis is not representative, since it has been condensed for didactic purposes.}
    \label{moment}
\end{figure}

A moment-ratio plot is a graph in which a distribution is represented as a pair of standardized moments plotted on a single set of coordinate axes \cite{Vargo}. Introduced by \cite{Craig}, and further developed in \cite{Joko}, they represent an interesting way of visualizing distributions, and of discriminating among them. Some distributions may be represented as a set of points, some others as curves, and in the case of generalized distributions and families of distributions as areas. \\
Surprisingly, in the econophysical literature (and more in general in the recent statistical papers about the distributional properties of many economic quantities), moment-ratio plots are somehow neglected. Our aim is to show how they can be efficiently used to complement the information provided by other more famous plots, such as the Zipf and the meplot ones.\\
The typical standardized moments involved in moment-ratio plots are the coefficient of variation
\begin{equation}
CV=\gamma_2=\frac{\sigma_X}{\mu_X},
\end{equation}
the skewness
\begin{equation}
\gamma_3=E\left[\left(\frac{X-\mu_X}{\sigma_X} \right)^3 \right],
\end{equation}
and the kurtosis
\begin{equation}
\gamma_4=E\left[\left(\frac{X-\mu_X}{\sigma_X} \right)^4 \right].
\end{equation}
However standardized moments of higher order can be used as well. We refer to \cite{Vargo} for more details.\\
After investigating different possible alternatives, we have come to the conclusion that the best moment-ratio plot for the typical (size) distributions arising in econophysics is a simpler version of the CV-Skewness diagram of \cite{Vargo}. In this graph the information related to a given distribution is summarized by the behavior of the pairs of CV and skewness. In particular, for our purposes, it is sufficient to focus our attention on the distributions lying in the first quadrant.\\
An immediate consequence of the choice of a CV-Skewness moment-ratio plot is the following: Pareto distributions can only be represented for $\alpha>2$, since otherwise the variance does not exists, and hence the CV is not (theoretically) computable. Anyway, this is not a major problem, because even when the Pareto distribution is ruled out for $\alpha\leq 2$, all other distributions of interests, and especially the lognormal, are still treatable. \\
Figure \ref{moment} shows an example of discriminant moment-ratio plot for size distributions. We call it discriminant because we will use it to discriminate among possible candidate distributions. The picture is just for didactic purposes.\\
In this plot, all distributions that are symmetric about the mean have skewness equal to 0. Moreover, since the CV can always be adjusted to take any value, by acting on the location and scale parameters, we find out that all symmetric distributions, such as the normal, the uniform and the Student-t are represented by the dotted line $\gamma_3=0$.\\
The plot is then split into 4 areas:
\begin{itemize}
\item The Paretian zone. This area is delimited from above by the theoretical Paretian CV-Skewness curve. This curve is only attained by Pareto I distributed random variables and is given by the couples
\begin{equation}
\gamma_2=\frac{1}{\sqrt{p(p-2)}}, \qquad \gamma_3=\frac{1+p}{p-3}\frac{2}{\sqrt{1-2/p}},\qquad p>3.
\end{equation}
Notice that this curve has a limit point in $(0,2)$.\\
From below the Paretian zone is bounded by the inverted gamma distribution, represented by $\gamma_3=\frac{4\gamma_2}{1-\gamma_2^2}$ for $\gamma_2 \in (0,1)$.
\item The Gray zone. The Gray zone is delimited by the inverted gamma from above and by the lognormal from below. The lognormal curve is represented by the couples
\begin{equation}
\gamma_2=\sqrt{\omega-1}, \qquad \gamma_3=(\omega+2)\sqrt{\omega-1},
\end{equation}
where $\omega=\exp(\sigma^2)$. In Figure \ref{moment}, the lognormal CV-Skewness curve is shown as a dashed line.
\item The Lognormal zone. This area is constrained by the lognormal curve from above and by the gamma distribution from below. The latter is given by $\gamma_3=2\gamma_2$.
\item The Exponential/Thin Tailed zone. This is the zone below the gamma curve and above the Bernoulli one, $\gamma_3=\gamma_2-\frac{1}{\gamma_2}$.
\end{itemize}	
Now, imagine we have a data set with $X_1,...,X_n$ i.i.d. observations. We can easily compute the quantities
\begin{eqnarray}
\hat{\gamma}_2=\frac{\bar{X}}{\hat{\sigma}_X}=\frac{\frac{1}{n}\sum_{i=1}^nX_i}{\sqrt{\frac{1}{n-1}\sum_{i=1}^n (X_i-\bar{X})^2}}\\
\hat{\gamma_3}=\frac{1}{n}\sum_{i=1}^n\left(\frac{X_i-\bar{X}}{\hat{\sigma}_X} \right)^3.
\end{eqnarray}
The couple $(\hat{\gamma}_2,\hat{\gamma}_3)$ will then define a point in the discriminant moment-ratio plot of CV and skewness. The location of the point with respect to the four areas and the curves gives us a good idea of the possible candidate distribution. But let us see in more details:
\begin{itemize}
\item If our point falls in the Paretian zone, the distribution is likely to be of Paretian type. In particular Pareto I for points that lie on (or very close to) the Paretian curve. The more a point moves from the Paretian curve toward the inverted gamma one, the more likely the underlying distribution is not a Pareto I, but rather a Pareto II, a Fisk, or a general power law with a slowly varying component.
\item Similarly, if the point falls in the lognormal-like zone, the underlying distribution is likely to be lognormal-like. The closer the point is to the lognormal curve, the more likely the data will be lognormal. If the points falls in the lognormal-like zone, but more close to the gamma curve, then the data are likely to be closer to the generalized gamma family discussed in \cite{Klei}.
\item If the couple $(\hat{\gamma}_2,\hat{\gamma}_3)$ falls in the Exponential / Thin Tailed zone, both the lognormal and the Pareto are completely ruled out. Possible size distributions here are the Weibull and its generalizations or special cases.
\item In the case in which the point falls in the so-called Gray zone, more analyses are needed, since the discriminant moment-ratio plot is not able to give a totally reliable indication. Typically this area concerns mixtures of lognormal and power tails, lognormals with extremely large variances, and hybrid distributions such as the Yule one \cite{Joko}, \cite{Klei}. The Gray zone is also often visited in case of just a few observations in the data set.\\
Simulation studies allow to define the following rule of thumb: for values of CV smaller than 2, if the skewness is greater than 14, then the distribution is likely to be Paretian with good approximation, even if it falls within the Gray zone.
\item A point falling out of the four areas may represent a symmetrical distribution if it lies close to the dotted normal curve, or a mixture thin tailed distribution if it falls below the Bernoulli curve. However, since these are not cases of interest for us, we do not enter into much detail here. 
\end{itemize}	

\begin{figure}
  \centering
    \includegraphics[scale=0.7]{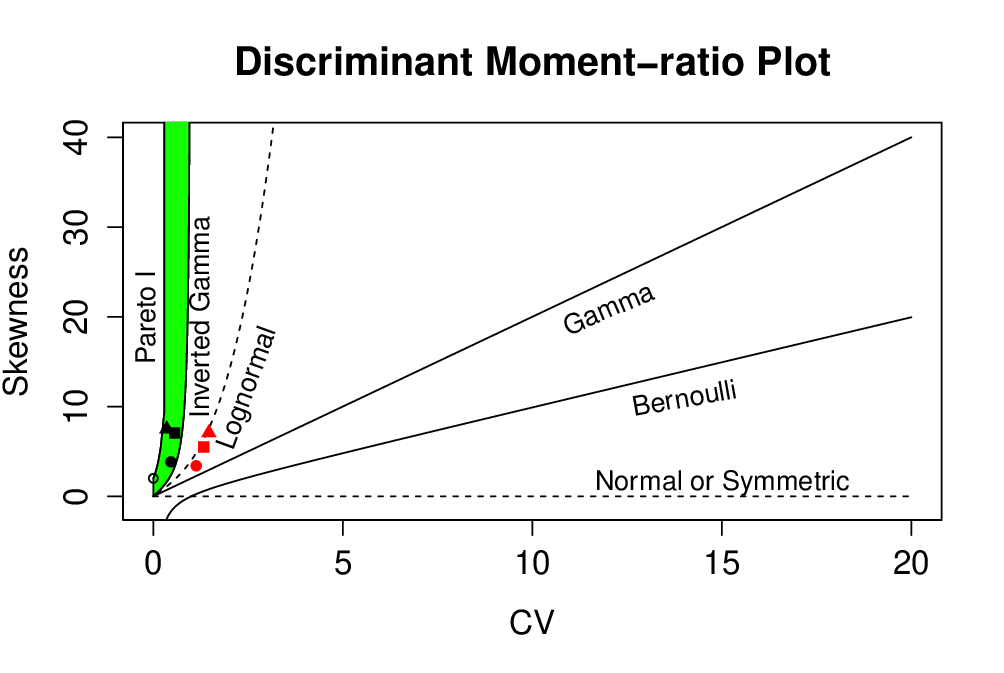}
  \caption{Application of the discriminant moment-ratio plot for the lognormal data of Figures 1 and 2 ($Lognormal(0,1)$). The red dot represents the couple $(\hat{\gamma}_2,\hat{\gamma}_3)$ for those data. The plot also shows how the location of the point changes if the size of the data set increases, from 500 (red dot) to 1000 (red square) and 5000 (red triangle). Black symbols show the points for a $Par(10,2.5)$ again for n=500 (black dot), 1000 (black square) and 5000 (black triangle).}
    \label{discri}
\end{figure}

The reliability of the discriminant moment-ratio plot increases with the number of observations (and obviously with the experience of the researcher). Good results are already obtainable with 100 or more observations. If the cardinality of the data sets is greater than 1000, the discrimination is strongly reliable. Simulation studies show that with 1000 observations the type one error for Pareto and Lognormal distributions is around $4\%$, decreasing to $1\%$ with more than 5000 observations. Similar results hold for the type two error. \\
If the size of the data set is particularly limited, a good idea can be to bootstrap the data and compute the couple $(\hat{\gamma}_2,\hat{\gamma}_3)$ for each sample. At this point it is possible to define the dispersion around the original point. We refer to \cite{Joko} and \cite{Vargo} for more details.\\
Figure \ref{discri} shows an application of the discriminant moment-ratio plot for the lognormal data we have already considered ($Lognormal(0,1)$). The red dot represents the couple $(\hat{\gamma}_2,\hat{\gamma}_3)$ for those data. Good news: the point clearly falls in the lognormal area and it is also fairly close to the lognormal curve. Differently from the Zipf and the mean excess plot, the discriminant moment-ratio one is able to clearly identify the lognormal nature of the data.\\
In the same plot we show how the location of the point changes if the size of the data set increases, from 500 (red dot) to 1000 (red square) and 5000 (red triangle). Black symbols show the points for a $Par(10,2.5)$ again for n=500 (dot), 1000 (square) and 5000 (triangle). The plot once again demonstrates a good discriminant power.\\
A simple R code to generate the discriminant moment-ratio plot is given in the appendix. The chosen configuration for the axes should cover most cases, however the code can be easily modified.

\section{The Zenga plot}\label{Zng}
\begin{figure}
  \centering
    \includegraphics[scale=0.6]{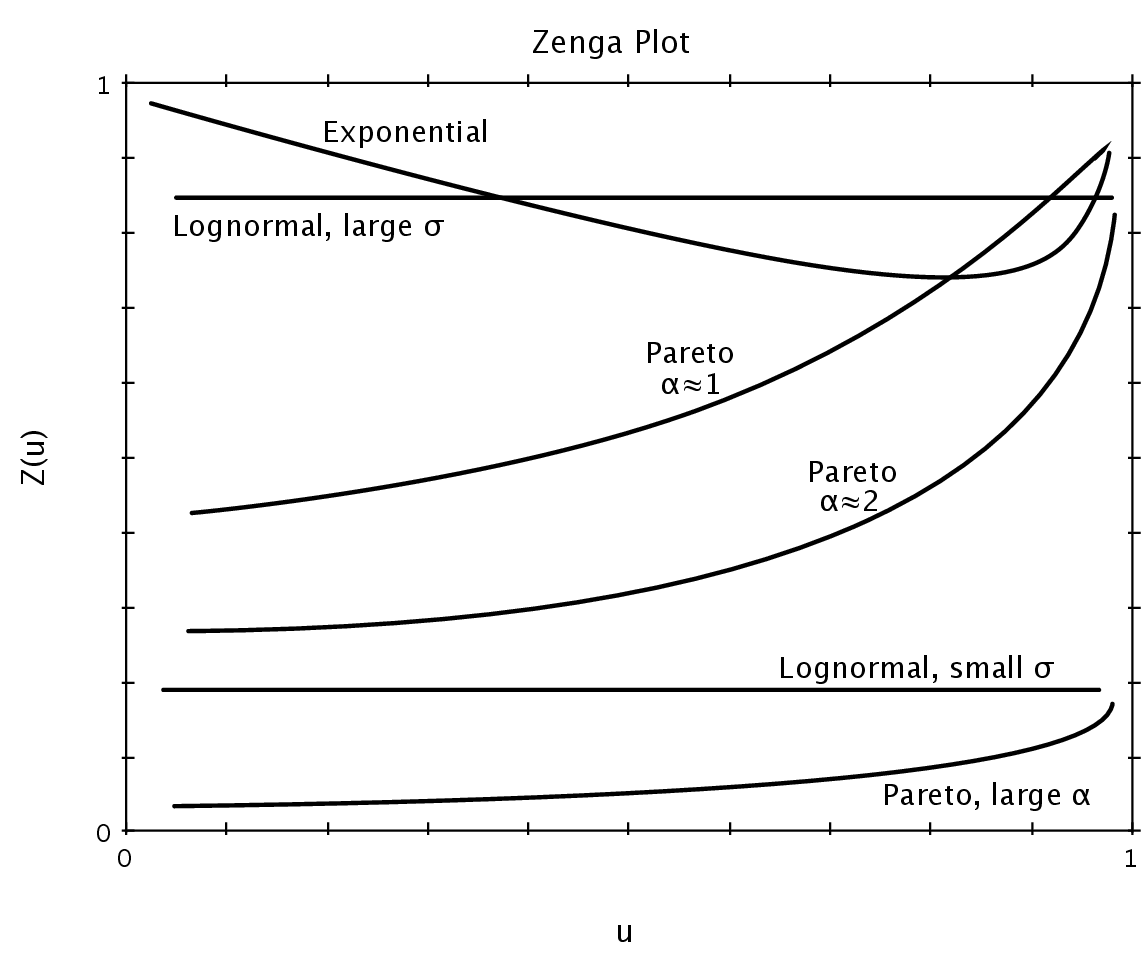}
  \caption{Zenga curve behavior for some classical size distributions.}
    \label{fig7}
\end{figure}

\begin{figure}
  \centering
    \includegraphics[scale=0.6]{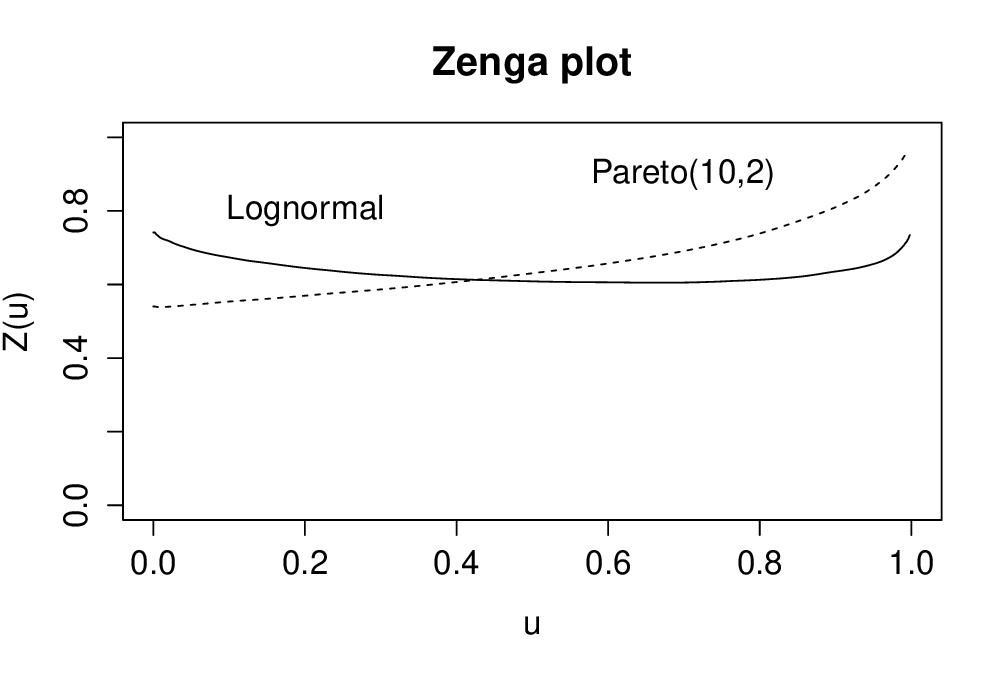}
  \caption{Zenga plot for the same standard lognormal data of Figures 1 and 2 (continuous line). The dashed line shows the Zenga curve of a sample of 500 observations from a Pareto(10,2).}
    \label{fig8}
\end{figure}

To our knowledge, the plot we propose in this section has never been used before to discriminate among possible size distributions for data. We call it Zenga plot, since it is based on the so-called Zenga curve, as presented in \cite{Zenga}.\\
The Zenga curve (see also \cite{Klei}, \cite{Zenga2}) represents an alternative to the well-known Lorenz curve as a measure of concentration. Since it is defined through the first-moment distribution, it exists only for $E[X]< \infty$. This implies that in the Paretian case, the Zenga curve is defined only for $\alpha>1$. Again this is not a great limitation: first of all $\alpha>1$ is something always observed in nature \cite{Clau}; moreover, even if the Pareto is ruled out, for $\alpha<1$, its competitors are still available.\\
Let $X$ be a nonnegative continuous random variable with support $(a,b)$, where $a$ and $b$ can be finite or infinite, density function $f(x)$, and distribution function $F(x)$. Let $\mu_x=E[X]<\infty$. \\
We can define the inferior mean and the superior mean as
\begin{equation}
\mu^-_x=\frac{1}{F(x)}\int_{a}^x s f(s) \text{d}s
\end{equation}
and
\begin{equation}
\mu^+_x=\frac{1}{1-F(x)}\int_{x}^b s f(s) \text{d}s
\end{equation}
respectively.\\
Now, by setting $x_{(u)}=F^{-1}(u)$ for $0<u<1$, we get
\begin{eqnarray}
Q^-_{(u)}=\mu^-_{F^{-1}(u)}=\frac{1}{u}\int_{0}^u x_{(s)} \text{d}s\\
Q^+_{(u)}=\mu^+_{F^{-1}(u)}=\frac{1}{1-u}\int_{p}^1 x_{(s)} \text{d}s.
\end{eqnarray}
The Zenga curve is hence given by
\begin{equation}
Z(u)=1-\frac{Q^-_{(u)}}{Q^+_{(u)}}, \qquad 0<u<1.
\end{equation}
As known, the Lorenz curve (e.g. \cite{Klei}) is defined as
\begin{equation}
L(u)=\frac{1}{E[X]}\int_0^u F^{-1}(s) \text{d}s, u\in[0,1].
\end{equation}
As a consequence, the Zenga curve can always be expressed via the Lorenz one, i.e.
\begin{equation}\label{loze}
Z(u)=\frac{u-L(u)}{u[1-L(u)]}, \qquad 0<u<1.
\end{equation}
Equation (\ref{loze}) is very important, because it allows us to derive the analytical form of the Zenga curve for many size distributions by simply plugging in the corresponding Lorenz curve. However, differently from the Lorenz curve, the Zenga one assumes a rather different shape for the diverse distributions we may be interested in, hence it represents a very useful tool to graphically discriminate them.\\
A classical Pareto I distribution with $F(x)=1-\left(\frac{x}{x_0} \right)^{-\alpha}$ for $0<x_0\leq x$ and $\alpha>1$, has $L(u)=1-(1-u)^{1-\frac{1}{\alpha}}$ and
\begin{equation}
Z(u)=1-(1-u)^{\frac{1}{\alpha(\alpha-1)}}.
\end{equation}
The Zenga curve of a Pareto distribution (and in general of Paretian distributions) is thus a convex increasing function on [0,1], and it approaches the $u$ axis for large values of $\alpha$, indicating a decrease in concentration.\\
In case of lognormally distributed random variables, the Zenga curve is constant and equal to
\begin{equation}
Z(u)=1-e^{-\sigma^2},\qquad 0<u<1,
\end{equation}
so that inequality increases with the variance.\\
For the exponential distribution, with $F(x)=1-e^{\lambda x}$ and $x,\lambda>0$, the Zenga curve is $Z(u)=-\frac{\log(1-p)}{p(1-\log(1-p))}$. Interestingly this curve does not depend on the parameter $\lambda$ of the underlying exponential distribution. It is convex with a minimum at $u=0.8336$.\\
Deriving the Zenga curve for any other distribution is quite simple. It is sufficient to use equation (\ref{loze}) and the functional form of the Lorenz curve of the desired distribution. For the explicit analytical forms of many Lorenz curves we refer to \cite{Sarabia} and \cite{Klei}.\\
In Figure \ref{fig7} the theoretical behavior of the Zenga curve for the Pareto, the Lognormal and the Exponential distributions is graphically given. It is clear why the Zenga plot can be a very good way to discriminate between, for instance, the lognormal and the Pareto distributions. While the Pareto always shows an increasing curve, the lognormal is constant.\\
The Zenga plot is rather easy to read and interpret; ambiguous cases are rarely observed. In the comparison between the lognormal and the Pareto distributions, for example, problems can rise when there is the need to choose between a lognormal with a very small standard deviation (e.g. $\sigma\leq0.5$) and a Pareto with an extremely large $\alpha$ (e.g. $\alpha\geq15$). But these limiting cases are definitely not observable when studying economic phenomena, such as the size distribution of companies, or the distribution of financial quantities, i.e. the typical arguments of econophysical investigation.\\
Let us once again apply the new plot to the same lognormal data set we have considered so far. Figure \ref{fig8} clearly shows how the empirical Zenga curve (black continuous line) is on average constant (and around $0.63=1-\exp(-1))$, apart from the two curvatures at the extremities\footnote{These curvatures depend on the empirical computation of the Zenga curve and they tend to become less and less relevant as the number of observations increases.}, suggesting the presence of lognormal data. In the same plot, for the reader's convenience, the empirical Zenga curve of a data set with $500$ observations from a $Par(10,2)$ is also given. The difference is clear.\\
At this point the question is: how to empirically compute the Zenga curve? The answer is twofold.\\ 
The easiest method is to compute the empirical Lorenz curve and to apply equation (\ref{loze}) in order to obtain the empirical Zenga. For the computation of the empirical Lorenz we refer to any good book in statistics. For a more specific and complete treatment, we suggest \cite{Choti}.\\
The second method requires to compute $\hat{Q}^-_{(u)}$ and $\hat{Q}^+_{(u)}$ first, and then $\hat{Z}(u)$.\\
Consider $N$ observations with $s\leq N$ distinct values $0\leq x_1 \leq ... \leq x_j \leq ... \leq x_s$, with frequencies $n_j$, $j=1,2,...,s$. For every $j$, define
\begin{eqnarray}
N_j=\sum_{i=1}^j n_i\\
u_j=N_j/N\\
T_j=\sum_{i=1}^j x_in_i\\
T=\sum_{j=1}^s x_jn_j.
\end{eqnarray}
Hence we have
\begin{equation}
\hat{Q}^-_{(u_j)}=\frac{T_j}{N_j},\qquad j=1,2,...,s,
\end{equation}
and
\begin{equation}
\hat{Q}^+_{(u_j)}=\begin{cases}\frac{T-T_j}{N-N_j}& j=1,2,...,s-1\\
x_s & j=s
\end{cases}.
\end{equation}
The empirical Zenga curve is thus obtained as $\hat{Z}(u_j)=1-\frac{\hat{Q}^-_{(u_j)}}{\hat{Q}^+_{(u_j)}}$.\\
As usual a simple code to generate the Zenga plot is provided in the appendix. 

\section{Conclusions}
The plots we have discussed so far do not represent the entire list of graphical tools one could use to check for Paretianity in the data.\\
A very common tool, especially among extreme value analysts, is the QQ-plot. As known, in a QQ-plot a distributional hypothesis is tested by plotting the empirical quantiles of the data against the theoretical quantiles of a candidate distribution. If the points in the plot more or less lie on the line $y=x$, then the empirical data are likely to come from the theoretical distribution we have chosen. Departures from this linear behavior generally indicate that the candidate distribution is not the correct one. To check for Paretianity, one can choose the Pareto distribution (or a similar power law) as the theoretical distribution to be checked. However, this is not the most common way. Most of the time, the exponential distribution is used as a benchmark, and Paretianity is signaled when the empirical data show fatter tails. Another possibility is to use the GEV, or generalized extreme value distribution \cite{Falk}, as the benchmark. We refer to \cite{Embr} for more details on QQ-plots, and to the strictly related probability plots, for power laws.\\
In \cite{Cihu} two additional plots are discussed. The first one relies on the scalability of sums for power laws. This property implies that Paretian random variables maintain their Paretian behavior even after aggregation. In particular, if we have a sample $X_1, X_2,...,X_n$ from a Pareto distribution, and we generate a new sample $X_1+X_2, X_3+X_4,...,X_{n-1}+X_n$, then the new sample will be still Pareto distributed with the same shape parameter, while the scale naturally changes. This property can be graphically tested by comparing the Zipf plots of the original data and of the aggregated ones. If the lines in the graph are more or less parallel, then this could be seen as a sign of power law behavior. We refer to \cite{Cihu} for a more complete discussion.\\
The other plot is a graphical representation of the extreme value test of \cite{Huli}. This test is based on the observation that power laws are in the domain of attraction of the generalized extreme value distribution \cite{Falk}. As a consequence, a random sample showing a power law tail must satisfy the extreme value conditions about the normalization of maxima. From a graphical point of view, one must study the behavior of the so-called $E_n$ statistics with respect to the empirical quantiles of the data of interest. We refer to \cite{Cihu} for more details.\\
Naturally, one could even introduce new tools. For example, it is known that the order statistics of a Pareto distribution have some interesting properties. A peculiar one is related to the so-called geometric spacings characterization \cite{Joko}. Consider a sample $X_1, X_2,...,X_n$. The sample comes from a Pareto distribution if the quantities $X_{i:n}$ and $\frac{X_{i+1:n}}{X_{i:n}}$, $i=1,...,n-1$, are independent (where $X_{i:n}$ is the $i$-th order statistic). Graphically, this can be verified using a simple scatter plot in which no particular dependence structure is observed.\\
However, all these additional tests represent, in our opinion, a refinement of the four ones we have discussed in the paper. In empirical analyses, they also show to be more difficult to be interpreted. Furthermore, the Zipf plot, the mean excess plot, the Zenga plot and the discriminant moment-ratio plot do allow for the simultaneous comparison of many different distributions at a time, while, say, a QQ-plot can only show if the empirical data come from the candidate distribution or not. For these reasons, we believe that the plots we have analyzed in the paper constitute the best options for the graphical, exploratory analysis of data. As said, these plots need to be combined, since each of them, even if with different levels of confidence, is not at all sufficient to have a good understanding of data.\\
It goes without saying that the graphical testing of the Paretian (or whatever kind of distributional) hypothesis is just the first step for a correct analysis. Only the combination of graphical and statistical tests can guarantee the desired level of confidence in studying actual data. Among the statistical tests, if the critical values are available, as they generally are in the Paretian class (see for example \cite{Arshad}), our personal preference goes to the Anderson-Darling one, since it better performs on tails. But this discussion would need another paper, and we will go into more details in the future.

\section*{Appendix: Codes}
In this appendix we collect some simple R codes that can be used to generate all the different plots we have discussed in the paper. These are the actually used programmes.\\
Naturally all the codes can be improved; these examples are simply given for the reader's convenience.

\subsection*{Zipf plot}
Here our basic code to produce Zipf plots.

\begin{verbatim}
zipfplot=function (data,type='plot',title=T) {
	
# type should be equal to 'points' if you want to add the
# Zipf Plot to an existing graph
# With other strings or no string a new graph is created.
# If title is set to be F, the title of the plot is not given. 
# This can be useful when embedding the Zipf plot into other 
# plots.

  data <- sort(as.numeric(data)) #sorting data
  y <- 1 - ppoints(data) #computing 1-F(x)
  if (type=='points'){
    points(data, y, xlog=T, ylog=T, xlab = "x on log  scale", 
    ylab = "1-F(x) on log scale")}
  else{
  	 if (title==F) {plot(data, y, log='xy', xlab = "x on log  scale", 
  	 ylab = "1-F(x) on log scale")} 
  	 else {plot(data, y, log='xy', xlab = "x on log  scale", 
  	 ylab = "1-F(x) on log scale", main='Zipf Plot')}
  	 }}
\end{verbatim}

\subsection*{Meplot}
Here we present a basic code for the mean excess function plot. More options can be added with no effort.

\begin{verbatim}
meplot=function(data,cut=5) {
# In cut you can specify the number of maxima you want to exclude.
# The standard value is 5

data=sort(as.numeric(data));
n=length(data);

mex=c();

for (i in 1:n) {
    mex[i]=mean(data[data>data[i]])-data[i];
}
data_out=data[1:(n-cut)];
mex_out=mex[1:(n-cut)];
plot(data_out,mex_out,xlab='Threshold u', ylab='Mean Excess e(u)',
main='Mean Excess Plot (Meplot)')
}
\end{verbatim}

\subsection*{Discriminant Moment-ratio plot}
The code we provide is meant to cover most of the cases one can observe with economic data, especially when studying the size distribution of firms.\\
That is why we restrict our attention to the first quadrant.\\
However, if your data produce a couple $(\hat{\gamma}_2,\hat{\gamma}_3)$ that lies out to the plot, the code can be easily modified.

\begin{verbatim}
moment_plot=function(data){
  
  # "data" is a vector containing the sample data

  ##############################################
  ##############################################
  # CV and Skewness functions
  coefvar=function(data){  
    CV=sd(data)/mean(data)
    CV}
  skewness=function(data) {
    m_3 <- mean((data-mean(data))^3)
    skew <- m_3/(sd(data)^3)
    skew}
  ##############################################
  ##############################################
  # Computation of CV and Skewness
  # CV
  CV=coefvar(data);
  # Skewness
  skew=skewness(data)
  # Rule of Thumb
  if (CV<0 | skew <0.15){print('Possibly neither Pareto 
                               nor lognormal. Thin tails.'); stop}

  ##############################################
  # Preparation of the plot
  ##############################################
  # Paretian Area
  # The upper limit - Pareto I 
  p=seq(3.001,400,length.out=250)
  g2brup=1/(sqrt(p*(p-2)))
  g3brup=(1+p)/(p-3)*2/(sqrt(1-2/p))
  # The lower limit, corresponding to the Inverted Gamma
  g2ibup=seq(0.001,0.999,length.out=250)
  g3ibup=4*g2ibup/(1-g2ibup^2)
  ##############################################
  # Lognormal area
  # Upper limit: Lognormal
  w=seq(1.01,20,length.out=250)
  g2log=sqrt(w-1)
  g3log=(w+2)*sqrt(w-1)
  # Lower limit - Gamma
  g2iblow=seq(0,20,length.out=250)
  g3iblow=2*g2iblow
  ##############################################
  # Exponential Area
  # The upper limit corresponds to the lower limit of the
  # lognormal area
  # The lower limit - Bernoulli
  g2below=seq(0,20,length.out=250)
  g3below=g2below-1/g2below
  ##############################################
  # The Gray area is obtained for free from
  # the previous lines of code.
  ##############################################
  # Normal / Symmetric distribution
  g2nor=seq(0,20,length.out=250)
  g3nor=rep(0,250)
  ##############################################
  # PLOT
  # Limits
  plot(g2iblow,g3iblow,'l',xlab='CV',ylab='Skewness',main='Discriminant 
  Moment-ratio Plot',xlim=c(0,20),ylim=c(-1,40))
  lines(g2ibup,g3ibup,'l')
  lines(g2brup,g3brup,'l')
  lines(g2below,g3below,'l') 
  lines(g2log,g3log,lty=2) # Lognormal
  lines(g2nor,g3nor,lty=2) # Normal
  # Strictly Paretian Area
  polygon(c(g2ibup,g2brup),c(g3ibup,g3brup),col='green')
  points(0,2,pch=1,cex=0.8) # Pareto limit point
  # Hints for interpretation
  text(-0.2,20,cex=0.8,srt=90,'Pareto I')
  text(1.2,20,cex=0.8,srt=90,'Inverted Gamma')
  text(2.5,12,cex=0.8,srt=70,'Lognormal')
  text(12,21,cex=0.8,srt=23,'Gamma')
  text(14,11,cex=0.8,srt=10,'Bernoulli')
  text(15,1.5,cex=0.8,'Normal or Symmetric')
  points(CV,skew,pch=16,col='red')
  return(c(CV,skew))
}
\end{verbatim}
\newpage
\subsection*{Zenga plot}
The code we provide makes use of the $Lc$ function of the $ineq$ package of R. An alternative code based on the procedure described in Section \ref{Zng} is easily implementable.

\begin{verbatim}
zengaplot=function(data){
# Since the code relies on the Lorenz curve
# as computed by the "ineq" library,
# we upload it 
library(ineq)
# Empirical Lorenz
est=Lc(data)
# Zenga curve
Zu=(est$p-est$L)/(est$p*(1-est$L))
# We rescale the first and the last point for 
# graphical reasons
Zu[1]=Zu[2]; Zu[length(Zu)]=Zu[(length(Zu)-1)]
# Here's the plot
plot(est$p,Zu,xlab='u',ylab='Z(u)',ylim=c(0,1), main='Zenga plot','l',lty=1)
}
\end{verbatim}
\end{document}